\documentclass[12pt,preprint]{aastex}

\begin{document}

\def\wisk#1{\ifmmode{#1}\else{$#1$}\fi}

\def\lt     {\wisk{<}}
\def\gt     {\wisk{>}}
\def\le     {\wisk{_<\atop^=}}
\def\ge     {\wisk{_>\atop^=}}
\def\lsim   {\wisk{_<\atop^{\sim}}}
\def\gsim   {\wisk{_>\atop^{\sim}}}
\def\kms    {\wisk{{\rm ~km~s^{-1}}}}
\def\Lsun   {\wisk{{\rm L_\odot}}}
\def\Zsun   {\wisk{{\rm Z_\odot}}}
\def\Msun   {\wisk{{\rm M_\odot}}}
\def\um     {$\mu$m}
\def\mic     {\mu{\rm m}}
\def\sig    {\wisk{\sigma}}
\def\etal   {{\sl et~al.\ }}
\def\eg     {{\it e.g.\ }}
 \def\ie     {{\it i.e.\ }}
\def\bsl    {\wisk{\backslash}}
\def\by     {\wisk{\times}}
\def\half {\wisk{\frac{1}{2}}}
\def\third {\wisk{\frac{1}{3}}}
\def\nwm2sr {\wisk{\rm nW/m^2/sr\ }}
\def\nw2m4sr {\wisk{\rm nW^2/m^4/sr\ }}

\title{On the nature of the sources of the cosmic infrared background.}

\author{
A. Kashlinsky\altaffilmark{1,2,4}, R. G. Arendt\altaffilmark{1,2},
J. Mather \altaffilmark{1,3}, S. H. Moseley \altaffilmark{1,3} }
\altaffiltext{1}{Observational Cosmology Laboratory, Code 665,
Goddard Space Flight Center, Greenbelt MD 20771}
\altaffiltext{2}{SSAI}\altaffiltext{3}{NASA}\altaffiltext{4}{e--mail:
kashlinsky@stars.gsfc.nasa.gov}

\begin{abstract}
We discuss interpretation of the cosmic infrared background (CIB)
anisotropies detected by us recently in the Spitzer IRAC based
measurements. The fluctuations are approximately isotropic on the
sky consistent with their cosmological origin. They remain after
removal of fairly faint intervening sources and must arise from a
population which has a strong CIB clustering component with only a
small shot-noise level. We discuss the constraints the data place
on the luminosities, epochs and mass-to-light ratios of the
indvidual sources producing them. Assuming the concordance
$\Lambda$CDM cosmology the measurements imply that the luminous
sources producing them lie at cosmic times $<$ 1 Gyr and were
individually much brighter per unit mass than the present stellar
populations.
\end{abstract}


\section{Introduction}

If the early Universe contained significantly more luminous
populations than at present, such as is thought to be the case
with the very first metal-free stars (see review by Bromm \&
Larson 2005), these populations could have produced a significant
contribution to the cosmic infrared background (CIB) with
potentially measurable structure (Santos et al 2002, Salvaterra \&
Ferrara 2003, Cooray et al 2004, Kashlinsky et al 2004; see
Kashlinsky (2005) for recent review). In an attempt to uncover the
CIB fluctuations from early populations we have analyzed deep
images obtained with the Spitzer IRAC instrument (Kashlinsky,
Arendt, Mather \& Moseley 2005; hereafter KAMM1), which led to
detecting significant CIB fluctuations remaining after subtracting
sources to faint flux levels. In a companion paper we presented
analysis from deeper and larger fields using the GOODS Spitzer
data (Kashlinsky, Arendt, Mather \& Moseley 2006; hereafter
KAMM2), which confirms our earlier findings and extends them to
fainter levels of removed galaxy populations and larger angular
scales.

In this {\it Letter} we discuss the cosmological implications of
the recent measurements of the CIB fluctuations from early
populations obtained by us (KAMM1,KAMM2). These measurements imply
that the signal must come from cosmic sources which have a
significant clustering component, but a low shot-noise
contribution to the power spectrum. Given the amplitude of the CIB
flux expected from these populations in the concordance
$\Lambda$CDM cosmology ($\gsim$1 \nwm2sr ), we show that these
sources must have very faint individual fluxes of $\lsim$10-20 nJy
in order not to exceed the measured levels of the remaining
shot-noise. Furthermore, these populations had to have
mass-to-light ratio significantly below that of the present day
stellar populations in order to produce the required CIB fluxes in
the short cosmic time available ($<$1 Gyr) from the available
baryons. Finally, we discuss the prospects for their individual
detection with future space missions. We use the AB magnitude
system, so flux per frequency $\nu$ of magnitude $m$ is
$S_\nu(m)$=$3631 \times 10^{-0.4m}$ Jy; diffuse flux in \nwm2sr is
defined as $\nu I_\nu$, with $I_\nu$ being the surface brightness
in MJy/sr.

\section{Magnitudes and epochs of the sources of the CIB fluctuations}

In their analysis KAMM1 and KAMM2 used a total of 5 different
fields with deep Spitzer IRAC observations of up to 24 hours per
pixel. All the observed fields are located at high Galactic and
Ecliptic latitudes and are free of significant zodiacal emissions
at all IRAC channels and of cirrus at the IRAC channels 1-3 (3.6
to 5.8 \um). Individual galaxies and other sources were removed
until a fixed level of the shot noise from remaining sources was
reached. The power spectrum of the remaining diffuse emission
showed a residual shot-noise component on small angles and a
significant excess due to clustering of faint/distant sources at
scales $\gsim 0.5^\prime$. Within the errors all fields cleaned to
the same shot-noise level showed the same excess fluctuations
consistent with their cosmological origin (see Fig. 1 of KAMM2).
At 8 \um\ there is a significant pollution by the Galactic cirrus
and at 5.8 \um\ the larger instrumental noise leads to relatively
large errors in the large-scale fluctuations. Here we concentrate
on the interpretation of the data at 3.6 and 4.5 \um\ in terms of
the luminosities, the epochs and the nature of the cosmological
sources contributing to these fluctuations.

KAMM1-2 show that the CIB fluctuations must come from cosmological
sources, such as ordinary galaxies and the putative Population
III. The former are defined as metal-rich stars with IMFs of a
Salpeter-Scalo \cite{kennicutt} type with masses $\sim 1M_\odot$.
Population III is defined (loosely) as luminous sources that
existed at, say, $z\gsim 10$ which possibly were individually very
massive and intrinsically very luminous. Data such as discussed
here cannot resolve whether the sources contributing to the CIB
were metal-rich \cite{komatsu} and whether the source of this
radiation was stellar nucleosynthesis \cite{santos} or black-hole
accretion in the early Universe \cite{pop3-qso}. Population III
epochs, $z\gsim$10, may contain emissions by both stars and
quasar-like objects \cite{kr83}.

\begin{figure}
\plotone{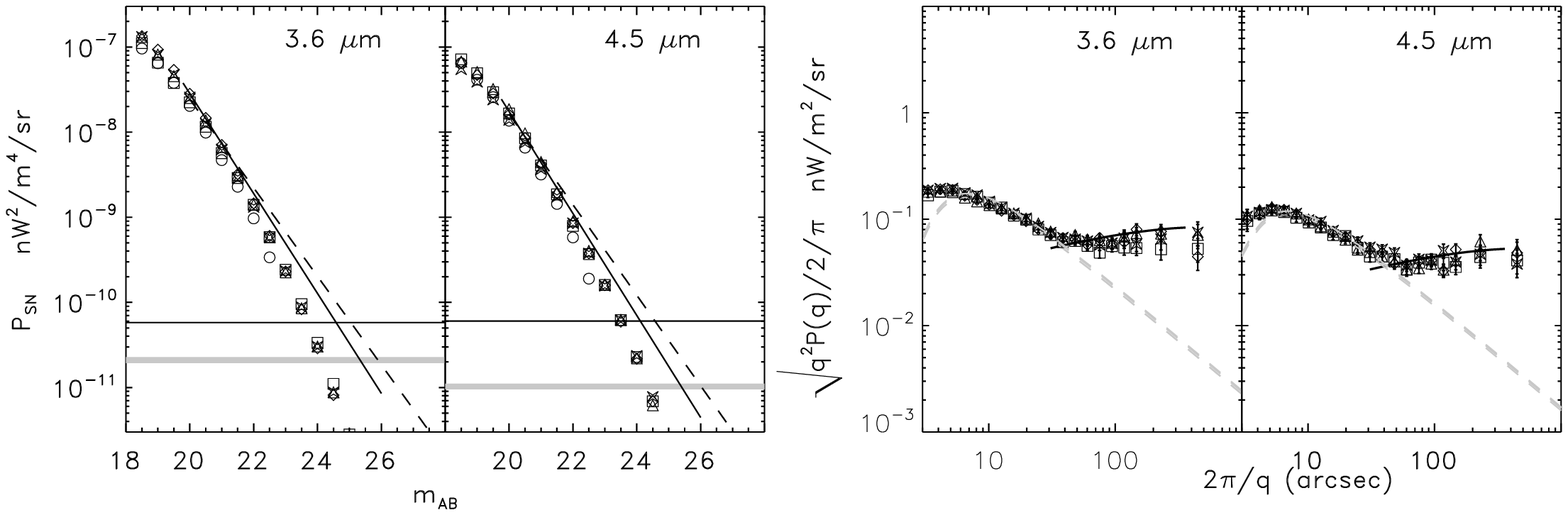}
\caption{Left: Shot noise power amplitude from the data is
compared to the values of $P_{\rm SN}$ estimated by integrating
the counts. Solid lines show the levels of $P_{\rm SN}$ reached in
the QSO1700 analysis \cite{kamm1}. Light shaded areas show the
levels of $P_{\rm SN}$ reached in KAMM2. Symbols plot $P_{\rm SN}$
by integrating the counts evaluated for all five fields in Table 1
of KAMM2. Diamonds correspond to HDFN-E1 region, triangles to
HDFN-E2, squares to CDFS-E1 and asterisks to CDFS-E2; open circles
correspond to counts for the QS1700 field. Solid line shows
$P_{\rm SN}$ according to the fit to IRAC counts of \cite{fazio}
used in \cite{kamm1}; dashed lines correspond to the IRAC counts
analysis from \cite{so}. The counts are significantly incomplete
due to confusion at the levels of $P_{\rm SN}$ reached with our
analysis and give a {\it lower} limit on the limiting magnitude.
Right: CIB fluctuations from KAMM2 at the shot-noise levels shown
with shaded regions in the left panel. The notations for the
counts from the GOODS data is the same as in the left panels.
Light-shaded dashes show the shot noise fluctuations. Solid lines
show the least squares fits to the CIB fluctuations from sources
at $z\gsim 6$ assuming $\Lambda$CDM model as described in the
text.  } \label{fig:shot-noise}
\end{figure}

Any model aimed to explain the CIB fluctuations results must
reproduce three major aspects: 1) The sources producing the
measured CIB fluctuations must be fainter than those removed from
the data. 2) They must reproduce the observed excess CIB
fluctuations at $\gsim 0.5^\prime$, where $\delta F
\simeq$0.07-0.1 \nwm2sr . 3) Lastly, the populations below the
above cutoff must account not only for the correlated part of the
CIB, but must also reproduce the observed (low) shot-noise
component of the signal. These lead to:

1) The the shot-noise component of the power spectrum from source
counts $dN/dm$ per magnitude interval $dm$ is $P_{\rm SN}$=$\int
S^2(m)dN(m)$ with $S$=$\nu S_\nu$. To estimate limiting magnitudes
implied by the measured shot-noise, we generated source counts for
the observed fields with SExtractor \cite{sextractor}. Fig.
\ref{fig:shot-noise} shows the remaining shot noise levels in
KAMM1-2 analysis and the counts data. The intersection of the
counts with the lowest shot noise levels shows that the sources
are eliminated to $m\!\gsim$25-26, so the detected CIB
fluctuations come from fainter sources. This magnitude limit at
3.6 \um\ corresponds to only $10^9 h^{-2}
10^{-0.4(m-25.5)}L_\odot$ emitted at 6000 \AA\ at $z$=5 where $h$
is the Hubble constant in units of 100 km/sec/Mpc. If the counts
contain extra populations in addition to those from \cite{fazio},
the magnitude limit will be fainter. Thus KAMM1,KAMM2 have removed
a significant fraction of galaxies even at $z$=5 and the CIB
fluctuations must come from sources at higher $z$.

2) The clustering component of the CIB at $ 0.5^\prime \lsim
2\pi/q \lsim 5^\prime$ requires $F_{\rm CIB}\sim$ a few \nwm2sr as
noted by us earlier \cite{kamm1}. The rms fluctuation in the CIB
flux, $\delta F$=$\sqrt{q^2P_2(q)/2\pi}$, on angular scale
$2\pi/q$ is related to the underlying 3-dimensional power spectrum
of the emitters' clustering, $P_3(k)$, the duration over which the
flux was produced, $\Delta t$, and the rate of the CIB production
rate, $dF/dt$, via the Limber equation (e.g. Kashlinsky 2005a):
\begin{equation}
\delta F = F_{\rm CIB} \bar{\Delta}_F  \;\; ; \;\;
\bar{\Delta}_F^2\equiv \frac{\Delta t \int_{\Delta t} (dF/dt)^2
\Delta^2(qd_A^{-1}) dt}{[\int_{\Delta t} (dF/dt) \; dt]^2}
\label{cib}
\end{equation}
where $\Delta(k)=[k^2P_3(k)/2\pi c\Delta t]^{1/2}$ is the rms
fluctuation in source counts over the cylinder of radius $2\pi/k$
and length $c\Delta t$. In the limit when the CIB release rate is
approximately constant, the relative CIB fluctuation,
$\bar{\Delta}_F$, will be $\sim\langle
\Delta^2(qd_A^{-1})\rangle^{1/2}$ with $\langle \ldots \rangle
\equiv (\Delta t)^{-1}\int_{\Delta t} \ldots dt$. If $dF/dt$ peaks
at some cosmic epoch $z_p$, the relative fluctuation will be
$\simeq \Delta(qd_A^{-1}(z_p))$.

To evaluate the range of the expected CIB flux from the sources
producing the measured fluctuations, we adopt the $\Lambda$CDM
model with $(\Omega,\Omega_{\rm baryon}, \Omega_\Lambda,
h)$=(0.3,0.044,0.7,0.71) and consider the epochs spanning 5$\leq\!
z \!\leq$20. The cosmic time at $z$=20 is $\simeq$0.2 Gyr and the
time between $z$=20 and $z$=5 is 1 Gyr. The scale
$r_8$=$8h^{-1}$Mpc, with today's density contrast $\sigma_8$,
subtends $\theta_8 \simeq$(3-4)$^\prime$. The relative fluctuation
in the projected 2-dimensional power spectrum, $\Delta$, on that
angular scale $\theta_8$, produced from sources located at mean
value of $\bar{z}$ and spanning the cosmic time $\Delta t$, would
be $\Delta(\theta_8)\sim \sigma_8 (1+\bar{z})^{-1}(r_8/c\Delta
t)^{1/2}\simeq 0.02 \sigma_8
(\frac{\bar{z}}{10})^{-1}(\frac{\Delta t}{{\rm Gyr}})^{-1/2}$
neglecting the amplification due to biasing. Biasing, due to
sources forming out of rare peaks of the density field, will
increase $\Delta$ \cite{kaiser} and for reasonable bias factors
($\sim 2$ for systems collapsing at $z\sim$5 to $\gsim 3$ at
$z\gsim 10$) one can gain amplification factors, $A$, in $\Delta$
of $\simeq 2$ to $\gsim$4-5 between $z$=5 and 20
\cite{k91,k98,cooray,kagmm}. Thus the arcminute scale CIB
fluctuations of $\delta F\sim$0.07-0.1 \nwm2sr at 3.6, 4.5 \um\
require the mean CIB from these sources to be $F_{\rm CIB} \sim
4-5 \langle A [(1+z)/6]^{-1}\rangle^{-1} (\Delta t/1{\rm
Gyr})^{1/2}$ \nwm2sr . Assuming that the fluctuations are produced
by low surface brightness systems at much lower $z$ does not alter
the required high value of their mean CIB contribution: e.g.
taking $\Delta t$=5Gyr corresponding to the cosmic time between
$z$=1 and 20 gives $\bar{\Delta}\simeq$0.02 at 1$^\prime$ assuming
no biasing. (As discussed below such sources would likely produce
shot-noise in excess of what we measure.) We can reach similar
conclusions with the entire range of scales $\gsim 0.5^\prime$
where we measure the clustering component of the CIB. The left
panels of Fig. \ref{fig:shot-noise} show the least squares fits
for $F_{\rm CIB}$, assuming the $\Lambda$CDM model, from all the
fields data at 3.6 and 4.5 \um. This gives $F_{\rm CIB} \langle A
(\frac{1+z}{10})\rangle \gsim (4, 2.5) (\Delta t/1{\rm
Gyr})^{-1/2}$ \nwm2sr at (3.6,4.5) \um\ respectively.

We thus conservatively take the fiducial flux of $F_{\rm CIB}$=1
\nwm2sr as the minimal CIB flux at 3.6 and 4.5 \um\ required by
the fluctuations, corresponding to the relative minimal CIB
fluctuations of $\sim$7\%. The results below can be re-scaled to
arbitrary $F_{\rm CIB}$, but our general conclusions will be valid
unless the CIB flux from sources producing the measured
fluctuations is significantly {\it below} the above number.
Although the net CIB fluxes may be, in principle, much higher,
this {\it minimal} CIB level at 3.6 \um\ is smaller than the
claimed CIB excess from DIRBE and IRTS measurements over that from
galaxy counts \cite{dwekarendt,arendtdwek,irts}, and is consistent
with the recent measurements of absorption in the spectra of
fairly distant ($z$=0.13-0.18) blazars at TeV energies
\cite{dwek05,hess}. Such CIB levels should, however, be measurable
from the spectra of gamma-ray bursts at $z\gsim$1-2 detectable
with the upcoming NASA's GLAST mission out to 300 Gev \cite{grbs}.
Spitzer counts \cite{fazio} show that the remaining ordinary
galaxies can contribute only $\simeq$0.15 \nwm2sr at 3.6 \um\
\cite{kamm1}, so to explain the CIB fluctuations with the
remaining (extrapolated) Spitzer counts sources requires almost
$\sim$100\% fluctuation on arcminute scales.

3) The CIB in the populations producing the measured fluctuations
significantly exceeds that from extrapolated IRAC counts
\cite{fazio}, so the excess flux must come from fainter
populations with a significant deviation from the extrapolated
counts slope \cite{kamm1}. The measured fluctuations indicate a
population with a relatively strong clustering component, which at
the same time has low shot noise. This means that the sources must
be individually faint. The shot-noise from the remaining galaxies
dominates the power spectrum of the CIB at $\lsim$0.5$^\prime$ and
its amplitude sets an {\it upper} limit on the shot-noise
component of the sources contributing to the arcminute scale CIB
fluctuations. The amplitude of the shot-noise component is
\cite{review}: $P_{\rm SN}$=$\int_{>m}S(m) dF(m)\!\equiv\!
S(\bar{m}) F_{\rm tot}(>\!m)$, where $dF(m)$=$S(m) dN(m)$ is the
CIB from sources at the magnitude interval $dm$ and $F_{\rm
tot}(m)$ is the total flux from the remaining sources of $>\!m$.
The sources contributing to the clustering component of the
fluctuations at arcminute scales must not exceed the level of the
residual shot noise in the data of $P_{\rm SN}\simeq (2,1) \times
10^{-11}$ nW$^2$/m$^4$/sr at (3.6, 4.5) \um. At 4.5 \um\ this
shot-noise amplitude of $P_{\rm SN}$=$10^{-11}$ nW$^2$/m$^4$/sr or
$10 (\lambda/3\mu{\rm m})^{-1} $ nJy$\cdot$\nwm2sr , would lead to
sources contributing to the signal having mean fluxes less than 12
$(F_{\rm CIB}/{\rm nWm^{-2}sr^{-1}})^{-1}$ nJy or AB magnitudes
$\bar{m} \geq 29 + 2.5 \lg(\frac{F_{\rm CIB}}{{\rm
nWm^{-2}sr^{-1}}})$. At 3.6 \um\ the shot-noise levels are a
factor of $\simeq 2$ larger leading to $\bar{m}$ about one
magnitude brighter. An important further information could be
obtained in still deeper measurements by setting a lower limit on
the shot-noise component of the sources contributing to the CIB
fluctuations determined when the clustering component disappears
of is substantially reduced.

\section{Discussion}

More information on the nature of populations of these faint
sources can be obtained by considering the fraction of baryons
that went through stars prior to $z\gsim 5$ ($\Delta t \lsim$1
Gyr) needed to explain the level of the CIB required by our data.
The net flux at frequency $\nu$ produced by the population with
comoving luminosity density ${\cal L}$, is $F_{\rm
CIB}$=$\frac{c}{4\pi}\int_{\Delta t} {\cal L}_{\nu^\prime}
(1+z)^{-1} dt$, where $\nu^\prime$=$\nu(1+z)$. This requires the
average comoving luminosity density at
(0.36-0.45)\um$\frac{10}{1+z}$ of:
\begin{equation}
{\bar{\cal L}} \simeq \frac{4\pi}{c}F_{\rm CIB}(\Delta t)^{-1}
(1+{\bar z}) \simeq 1.2\times 10^9 L_\odot{\rm Mpc}^{-3}\; \frac{1
{\rm Gyr}}{\Delta t} \frac{1+{\bar z}}{10} \frac{F_{\rm CIB}}{{\rm
nW m}^{-2}\mbox{sr}^{-1}} \label{eq:lumden}
\end{equation}
For comparison the present-day luminosity density measured by the
Sloan Digital Sky Survey at 0.32 \um\ to 0.68 \um\ is about an
order of magnitude lower \cite{blanton}. This indicates
significantly more luminous populations contributing to the CIB
fluctuations than at present. The contribution to the density
parameter by these sources is thus given by:
\begin{equation}
\Omega_{\rm *} = \frac{(\Gamma \; \bar{{\cal
L}})|_{(0.36-0.45)\mu{\rm m} \frac{10}{1+z}} }{3H_0^2/8\pi G}
\simeq
 8.3\times10^{-3} \frac{F_{\rm CIB}}{{\rm nW m^{-2}sr^{-1}}}
\;\frac{\Gamma}{\Gamma_\odot} \left(\frac{\Delta t}{1 {\rm Gyr}}
\right)^{-1} \frac{1+\bar{z}}{10}
 \label{eq:omega_lum}
\end{equation}
where $\Gamma$ is the mass-to-light ratio. For comparison the mean
density in present day stars is significantly lower at
$\Omega_{*,z=0}\! \simeq \!2\!\times\!10^{-3}$
\cite{fukugita,cole} and much of the contribution to
$\Omega_{*,z=0}$ comes from the late stellar Population I stars
with solar metallicities. Strictly speaking eq. \ref{eq:omega_lum}
assumes no re-processing of baryons and may thus overestimate the
required amount of luminous baryons in the case of short lived
massive stars, such as Population III, but it shows that it is
energetically easier to produce the significant CIB levels implied
by the Spitzer data in the cosmic time available with stars whose
mass-function is skewed toward $\Gamma \ll \Gamma_\odot$. (For
populations made up of massive stars it can be replaced with eq. 3
of Kashlinsky, 2005b). If the CIB fluctuations are produced by
populations containing a significant fraction of low-mass stars,
which should still be burning today, they would require a large
fraction of the present-day stars to have been produced at
$z\gsim$6-10.

\begin{figure}
\plotone{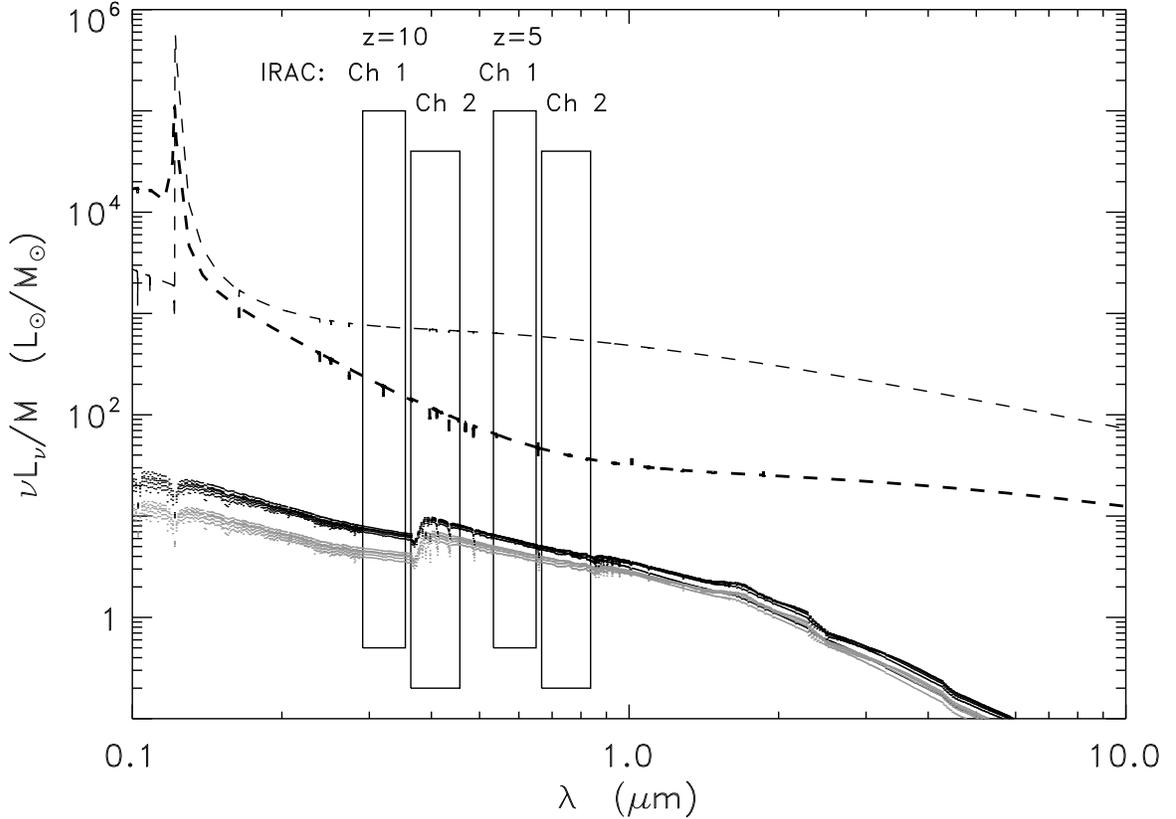}
\caption{Rest-frame luminosity per unit mass plotted vs
wavelength for Population III spectra (from Santos et al - dashed
lines) and ``ordinary" stellar populations at 0.5 and 1 Gyr with
Salpeter-Scalo IMF (computed from PEGASE for $Z=0, 10^{-3},2\times
10^{-3}, 5\times 10^{-3}, 10^{-1}$ assuming constant burst of star
formation, i.e. $SFR \propto \exp(-t/t_{\rm burst})$ with $t_{\rm
burst}=20$ Gyr.) $L_\odot = 3.8\times 10^{33}$ erg/sec is the
solar bolometric luminosity. The part of emissions probed by the
IRAC Channels 1 (3.6 \um) and 2 (4.5 \um) at $z$=5,10 is shown
with the marked regions.} \label{fig:m2l}
\end{figure}

To model the ordinary stellar populations, we have run stellar
evolution models using the PEGASE code \cite{pegase}, assuming
normal IMF with various metallicities and the ongoing star
formation (i.e. star formation rate $\propto \exp(-t/t_{\rm
burst})$ with $t_{\rm burst}$=20 Gyr). For Population III we
adopted the spectral energy distribution from \cite{santos}. Fig.
\ref{fig:m2l} shows the luminosity per unit mass in stars
($\Gamma^{-1}$) assuming the ordinary population to be less than 1
Gyr old ($\Gamma \sim$0.2-0.5$\Gamma_\odot$) and contrasts them
with the expectations for massive Pop III systems
($10^{-2}$-$10^{-3}\Gamma_\odot$). If the CIB fluctuation signal
comes entirely from the Population III systems, eq.
\ref{eq:omega_lum} would give the minimal fraction of baryons
locked in them $\sim 0.15 \% (F_{\rm CIB}/{\rm
nW\;m^{-2}sr^{-1}})$. If the baryons are re-used in stars this
fraction would be decreased. This number is in agreement with that
of \cite{grbs} after scaling to the appropriate CIB levels: 0.14\%
$(F_{\rm CIB, bolometric}/{\rm nW\;m^{-2}sr^{-1}}) (z/10)
(\epsilon/0.007)^{-1}$ assuming the hydrogen  burning efficiency
$\epsilon$. (Such massive stars would be fully convective with the
overall efficiency reaching $\epsilon \gsim 3\times 10^{-3}$,
Schaerer 2002).

The sources satisfying the above constraints had masses in
luminous matter of:
\begin{equation}
M_* \sim 4\pi d_L^2(1+z)^{-1} \;\Gamma S(\bar{m}) \lsim 7 \times
10^5 h^{-2} M_\odot \; \frac{\Gamma_{\frac{(3.6-4.5)\mu{\rm
m}}{(1+z)}}}{5\times 10^{-3}\Gamma_\odot}\;
\frac{[S(\bar{m})/\nu]}{20 {\rm nJy}} \;
\left(\frac{1+z}{10}\right)^{1.6} \label{mass}
\end{equation}
where the luminosity distance was approximated $d_L\simeq 3.2
(1+z)^{1.3}h^{-1}$Gpc. Such Population III sources, with only
$\lsim$ a few times $10^5 M_\odot$ in stellar material, would be
below the detection threshold in the high-$z$ Lyman dropout
searches of \cite{bouwens,willis} considered by \cite{sf06}. In
any case theoretical predictions of the luminosity function of
Population III sources are necessarily model-dependent as they
depend on the assumptions of the small-scale power and its
evolution as well as the microphysics governing the various
feedback effects during the collapse of the first haloes. The
Press-Schechter type prescriptions may break down for the slope
and regime of power spectra on the relevant scale \cite{springel}
and the feedback mechanisms due to the $H_2$ destruction by the
Lyman-Werner bands radiation \cite{haiman} likely suppress star
formation in a complicated halo-mass dependent way.

To resolve the faint sources responsible for the CIB fluctuations,
 their individual flux must
exceed the confusion limit usually taken to be $\alpha\geq 5$
times the flux dispersion produced by these emissions
\cite{condon}. Lower flux sources will be drowned in the confusion
noise; of course, this is precisely where CIB studies would take
off. If such sources were to contribute the CIB required by our
data, at 3.6 \um\ they had to have the average surface density of
$\bar{n} \sim F_{\rm CIB}^2/P_{\rm SN} \sim 2 \; {\rm arcsec}^{-2}
\; \left(\frac{F_{\rm CIB}}{{\rm nWm^{-2}sr^{-1}}}\right)^2
\left(\frac{P_{\rm SN}}{10^{-11}\; {\rm
nW^2m^{-4}sr^{-1}}}\right)^{-1}$. To avoid the confusion limit and
resolve these sources individually at, say, 5-sigma level
($\alpha$=5) one would need a beam area $\omega_{\rm beam} \leq
\alpha^{-2}/\bar{n} \sim 0.017\left(\frac{F_{\rm CIB}}{{\rm
nWm^{-2}sr^{-1}}}\right)^{-2}{\rm arcsec}^2 $
 or circular radius
$\lsim$0.07 $(F_{\rm CIB}/{\rm nWm^{-2}sr^{-1}})^{-1}$arcsec. This
is not in the realm of the current instruments, but the {\it JWST}
could be able to resolve these objects \cite{jwst}. Extrapolation
of this argument to shorter $\lambda$ is model-dependent as it
would assume both the SED of these sources (to predict their
magnitudes at $\lambda <$3 \um) and their $z$ (to predict the
location of their Lyman break and whether or not they are
observable at $\lambda <$3 \um). In any case, at 1.1 and 1.6 \um\
confusion is not reached until $m_{\rm AB} \gsim 28$
\cite{thompson}. If the first stars produced dusty environments
their far-IR luminosities will be substantial and these sources
should be visible at wavelengths redshifted today to mm and sub-mm
bands. In that case, they may be resolvable with the ALMA
\footnote{http://www.alma.nrao.edu/} large array, whose sub-mm
resolution is better than 0.02$^{\prime\prime}$.

Finally, the fluctuations are unlikely to come from low-luminosity
low-$z$ normal galaxies. Such galaxies must have the surface
density $\bar{n} \gsim 3\times 10^7$deg$^{-2}$ with the 3.6, 4.5
\um\ fluxes $\lsim$10-20 nJy. Unless they are significantly
fainter than this limit, emissions from star-forming systems
should have comparable flux at shorter $\lambda$ out to the 4000
\AA\ break for passively evolving populations or to the Lyman
cutoff at $\simeq$0.1 \um\ for star-forming galaxies. Galaxy
counts now extend to $m\simeq$30.5 (2  nJy) at 0.67 \um\ and to 29
(10 nJy) at 1.6 \um\ \cite{madaupozzetti} and are over an order of
magnitude below the required value of $\bar{n}$ at the faintest
magnitudes. This would exclude star forming galaxies as faint as 2
nJy at 0.67$/(1+z)$ \um\ at $z\lsim$5.7 and passively evolving
populations out to 10 nJy at 1.6$/(1+z)$ \um\ at $z\lsim$3. We
note, however, that this analysis cannot exclude ``abnormal"
populations at low $z$.

This work is supported by NSF AST-0406587 and NASA Spitzer
NM0710076 grants.

%\texttt{\{thebibliography\}}%

\clearpage

\clearpage

\end{document}